\documentclass[twocolumn]{aastex62}

\accepted{for publication in the ApJL September 1, 2019}

\shorttitle{Radial Dependence of Proto-GC Contribution to MW Formation}
\shortauthors{Chung et al.}
\begin{document}
\title{Radial Dependence of the Proto-Globular Cluster Contribution to the Milky Way Formation
}
\correspondingauthor{Chul Chung, Mario Pasquato}
\email{chulchung@yonsei.ac.kr, mario.pasquato@inaf.it}
\author[0000-0001-6812-4542]{Chul Chung}
\affil{Department of Astronomy \& Center for Galaxy Evolution Research, Yonsei University, Seoul 03722, Republic of Korea}
\author{Mario Pasquato}
\affiliation{INAF, Osservatorio Astronomico di Padova,
vicolo dell'Osservatorio 5, 35122 Padova, Italy}
\affiliation{INFN, Sezione di Padova,
Via Marzolo 8, I–35131, Padova, Italy}

\author[0000-0002-7957-3877]{Sang-Yoon Lee}
\affiliation{Department of Astronomy \& Center for Galaxy Evolution Research, Yonsei University, Seoul 03722, Republic of Korea}

\author{Ugo N. di Carlo}
\affiliation{INAF, Osservatorio Astronomico di Padova,
vicolo dell'Osservatorio 5, 35122 Padova, Italy}
\affiliation{INFN, Sezione di Padova, Via Marzolo 8, I–35131, Padova, Italy}
\affiliation{Dipartimento di Scienza e Alta Tecnologia, University of Insubria, Via Valleggio 11, I–22100, Como, Italy}

\author{Deokkeun An}
\affil{Department of Science Education, Ewha Womans University, 52 Ewhayeodae-gil, Seodaemun-gu, Seoul 03760, Republic of Korea}

\author[0000-0002-1842-4325]{Suk-Jin Yoon}
\affil{Department of Astronomy \& Center for Galaxy Evolution Research, Yonsei University, Seoul 03722, Republic of Korea}

\author[0000-0002-2210-1238]{Young-Wook Lee}
\affil{Department of Astronomy \& Center for Galaxy Evolution Research, Yonsei University, Seoul 03722, Republic of Korea}

\begin{abstract}

Recent interpretation of the color$-$magnitude diagrams of the Milky Way (MW) bulge has suggested that the observed double red-clump feature can be a natural consequence of He-enhanced stellar populations in the MW bulge. This implies that globular clusters (GCs), where the He-enhanced second-generation (SG) stars can be efficiently created, are the most likely candidate contributors of He-rich stars to the MW bulge. We extend this idea to the Galactic inner halo and investigate the fraction of the SG stars as a function of the Galactocentric distance. We use bluer blue-horizontal branch (bBHB) stars, which are assumed to be originated from He-rich SG populations, as proxies of SG stars, and find that the fraction of bBHB stars increases with decreasing Galactocentric distance. Simulations of the GC evolution in the MW tidal field qualitatively support the observed trend of bBHB enhancement in the inner halo. In these simulations, the increasing tidal force with decreasing Galactocentric distance leads to stripping of stars not only from the outskirts but also from the central regions of GCs, where SG stars are more abundant. We discuss the implication and prospect of our findings concerning the formation history of the bulge and inner halo of the MW.

\end{abstract}

\section{Introduction}
\label{introduction}

Understanding the formation history of the Milky Way is one of the ultimate goals of galaxy formation studies ~\citep[e.g.,][]{2015ARA&A..53...51S, 2014ARA&A..52..291C}. Various simulations of galaxy formation have revealed that, in addition to the stars that originated \emph{in-situ}, the most plausible candidates for the building blocks of the Milky Way are massive dwarf galaxies and\slash or massive globular clusters ~\citep[GCs; e.g.,][]{1991ASPC...13....3L}. Numerous studies on identifying the primary building blocks of the Milky Way have been carried out, but the nature of the building block that played a dominant role in building up the stellar populations of the Milky Way is still under debate. Especially the Galactic halo has been considered as a collection of smaller structures such as dwarf spheroidals, and it has been well established that dwarfs are the major constituents of the Milky Way halo via accretion ~\citep[e.g.,][]{1978ApJ...225..357S, 2007MNRAS.378..353K}. Compared to the dwarf galaxies, the contribution of GCs as building blocks of the Milky Way has been regarded as small because of their lower current mass without dark matter, as well as the small number of GCs~\citep{2009ARA&A..47..371T, 2015MNRAS.448L..77D}.

In this context, the discovery of He-enhanced populations in the Milky Way bulge, however, changed the current paradigms on the formation history of the Milky Way.
Proto-GCs{, where He- and N-enhanced stars can be formed \citep{2010A&A...519A..14M, 2016ApJ...825..146M, 2017MNRAS.465..501S},} are the most likely candidates for the birthplace of bulge He-enhanced stars through the self-enrichment of the first generation polluters ~\citep{2018ApJ...869...35K}, and the observed double red clumps (RCs) in the bulge (resulting from He-rich stars) are crucial evidence for their contribution to the bulge ~\citep{2015MNRAS.453.3906L, 2017ApJ...840...98J, 2019ApJ...878L...2L}.
The recent study on chemical tagging of the double RC in the bulge further strengthens this view by detecting {N-enhanced stars} in the bright RC ~\citep{2018ApJ...862L...8L}, which implies that almost 50\% of stars in the bulge may be originated from second-generation (SG)\footnote{Throughout this paper, the SG stars indicate all generations of stellar populations enriched by FG stars in proto-GCs.} stars.
This is in line with a picture of a rapid orbital decay of proto-GCs emerging from N-body models of Milky Way evolution, which suggests that the nuclear star cluster observed in the Galactic center is the result of repeated GC mergers \citep[e.g.,][]{2008MNRAS.388L..69C, 2018MNRAS.473..927A}.
Given the possibility of the significant fraction of GC-origin population in the bulge region, one would expect that the effect of proto-GCs on the formation of the halo should be non-negligible.

\citet{2016MNRAS.456L...1C} demonstrated the preferential removal of first-generation (FG) stars from GCs in the outer halo and cautiously suggested that this process may result in the misunderstanding that most of the outer halo consists of the dwarf-origin FG stellar populations.
On the other hand, the observed RC stars in the bulge ~\citep{2018ApJ...862L...8L} may be the simple outcome of full or almost full disruption of GCs at small Galactocentric distances.
Therefore, for intermediate Galactocentric distances, one would expect a partial stripping of SG stars, depending on the interplay between the strength of tidal stripping and the amount of confinement of the SG stars in the center of the GC, which in turn depends on the details of SG formation \citep[][]{2008MNRAS.391..825D, 2015MNRAS.450.1164H, 2018ARA&A..56...83B} and dynamical evolution within the host GC \citep[][]{2013MNRAS.429.1913V, 2014ApJ...791L...4D, 2016ApJ...832...99L}.
This may lead to a varying fraction between FG and SG stars in the bulge and halo fields.

In this study, we investigate how proto-GC systems, which usually have experienced multiple generations of star formation, have influenced the formation of the Milky Way inner halo. We analyze the number ratio between FG and SG stars as a function of the Galactocentric distance using the ratio of bluer blue horizontal branch (bBHB) to redder blue horizontal branch (rBHB) stars as proxies for each of the populations, respectively. We find a gradient of the bBHB$/$rBHB ratio with Galactocentric distance, which we interpret in terms of increasing preferential tidal stripping of FG stars as Galactocentric distance increases. This is expected in massive GCs, where relaxation did not have enough time to mix the different stellar generations and SG stars are strongly confined to the central regions of the GC \citep[e.g., see][]{2013MNRAS.429.1913V}.
To interprete our results, we additionally run simplified dynamical evolution simulations based on the numerical solution scheme for Hill equations proposed by \cite{2010AJ....139..803Q} for GCs at different Galactocentric distances.

The paper is organized as follows.
Section 2 describes the construction of a population synthesis model and the sample selection.
In Sections 3 and 4, we present our new findings together with results from simulations.
Section 5 discusses the implications of our results for the formation history of the Milky Way.

\section{Population Synthesis Model and Sample Selection}
\label{sampleselection}

The evolutionary population synthesis (EPS) models presented here are constructed under the same assumptions on the input parameters adopted in \citet{2016MNRAS.456L...1C}. Readers are referred to \citet{2013ApJS..204....3C} and \citet{2017ApJ...842...91C} for detailed prescriptions of the EPS models.

\begin{figure}
\centering
\includegraphics[angle=0, scale=0.8]{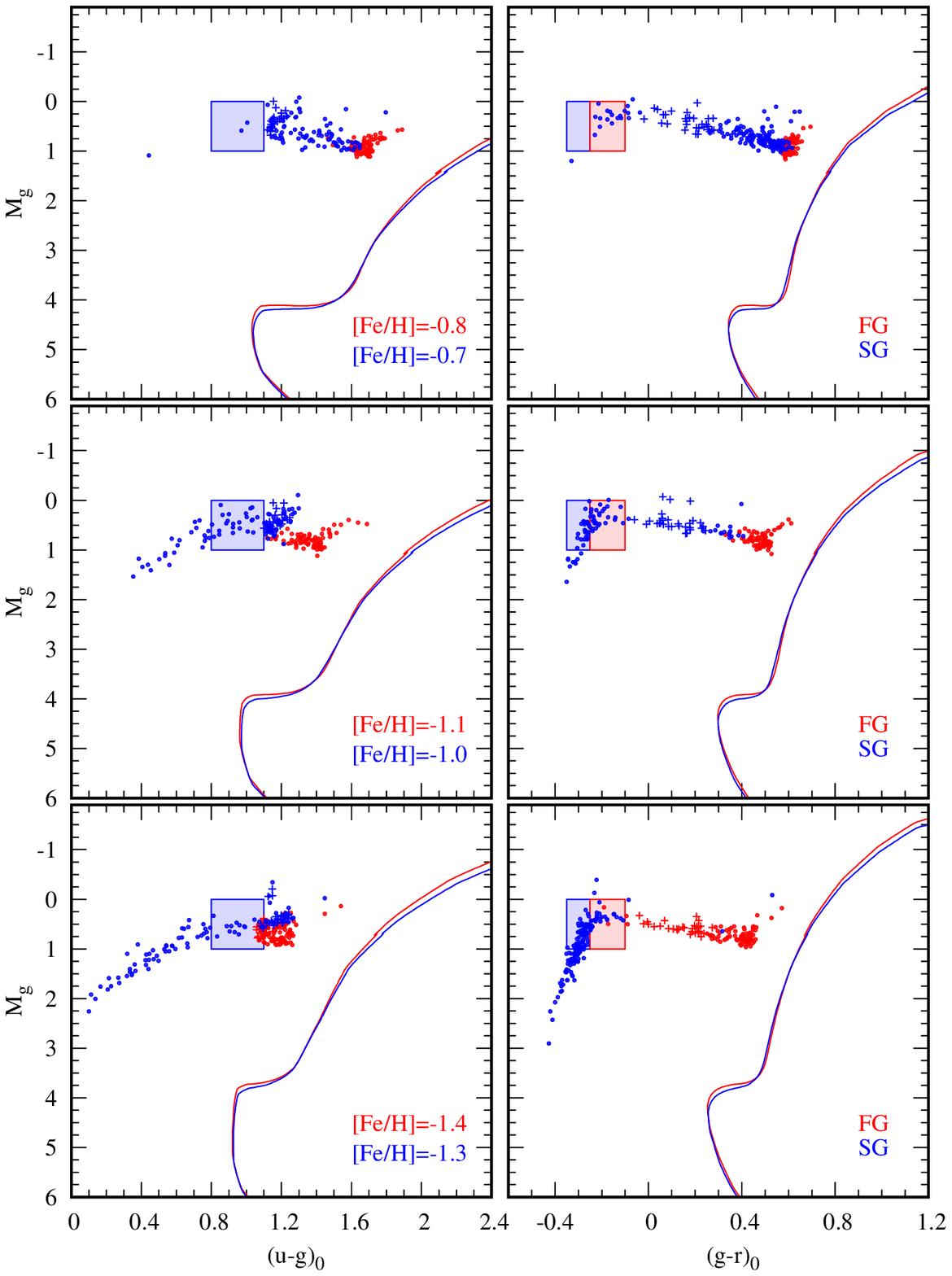}
\caption{The synthetic color$-$magnitude diagrams (CMDs) of proto-GCs with FG and SG populations.
The left and right panels are $(u-g)_0$ and $(g-r)_0$ CMDs at 12~Gyr, respectively.
Red and blue colors depict the FG and SG populations at given metallicities, respectively.
The metallicities of SGs are set as 0.1 dex higher than those of FGs.
Crosses between $-0.1 \leq (g-r)_0 \leq 0.3$ and $1.0 \leq (u-g)_0 \leq 1.3$ represent RR Lyrae stars.
Blue boxes indicate bBHB selection criteria at a given color, while red boxes in the right panels indicate where rBHBs are located in the CMDs.
\label{f1}}
\end{figure}

Figure~\ref{f1} presents synthetic horizontal branch (HB) stars and corresponding isochrones from the main sequence to the tip of red giant branch stars in $(u-g)_0$ and $(g-r)_0$ color--magnitude diagrams (CMDs) at given metallicities and 12~Gyr age.
Based on these CMDs, we carefully selected blue colored areas where only bBHBs are located in both colors of $0.8<(u-g)_0<1.1$ and $-0.3<(g-r)_0<-0.25$.
For rBHB stars, we adopted the selection criteria of ~\citet{2015MNRAS.448L..77D}.
To avoid RR Lyrae contamination in the sample, we took stars with $(g-r)_0 <-0.1$ as rBHB stars.
The mean absolute g-band magnitudes (${\rm M}_g$) of bBHBs within blue boxes in $(u-g)_0$ and $(g-r)_0$ colors are 0.49 and 0.70 {\emph mag}, respectively.
Given that the mean ${\rm M}_g$ of rBHBs in the blue box is 0.43 {\emph mag}, the ${\rm M}_g$ differences among selection boxes is less than 0.3 {\emph mag}, which can be used as distance indicator at a given magnitude.
From top to bottom panels, the metallicity of both FG and SG increases and bottom panels show the case of inner halo metallicity and age.
This metallicity range clearly shows the drastic morphology change of HB stars originated from SG populations at the age of 12 Gyr.

We make use of SDSS photometry ~\citep{2000AJ....120.1579Y} from DR14 for the halo star census.
Our candidate stars are selected from \emph{Photoprimary}.
We are mainly concerned with halo rBHB and bBHB stars at Galactocentric distances from 4 to 50~kpc, which corresponds to the apparent $g$ magnitude of rBHB stars ranging between 15 and 19 \emph{mag}.
We have trimmed the stars below $b = 30^\circ$ to avoid bulge or disk stars\footnote{The GAIA-Enceladus candidates exist everywhere in the halo but are well-known to have much more overlaps with the thick disk \citep[e.g.,][]{2018Natur.563...85H}. By selecting stars with $b>30^\circ$ and not including solar neighborhood stars, we have reduced the contamination caused by the GAIA-Enceladus.}, and selected our sample with photometric errors less than 0.05.
To {see the effect of the population shift in the same halo region} with respect to the Galactocentric distance, we choose a {rather narrow} volume-limited star sample in $-15^\circ < l < 15^\circ$ to meet the above criteria.

\begin{figure}
\centering
\includegraphics[angle=-90,scale=0.7]{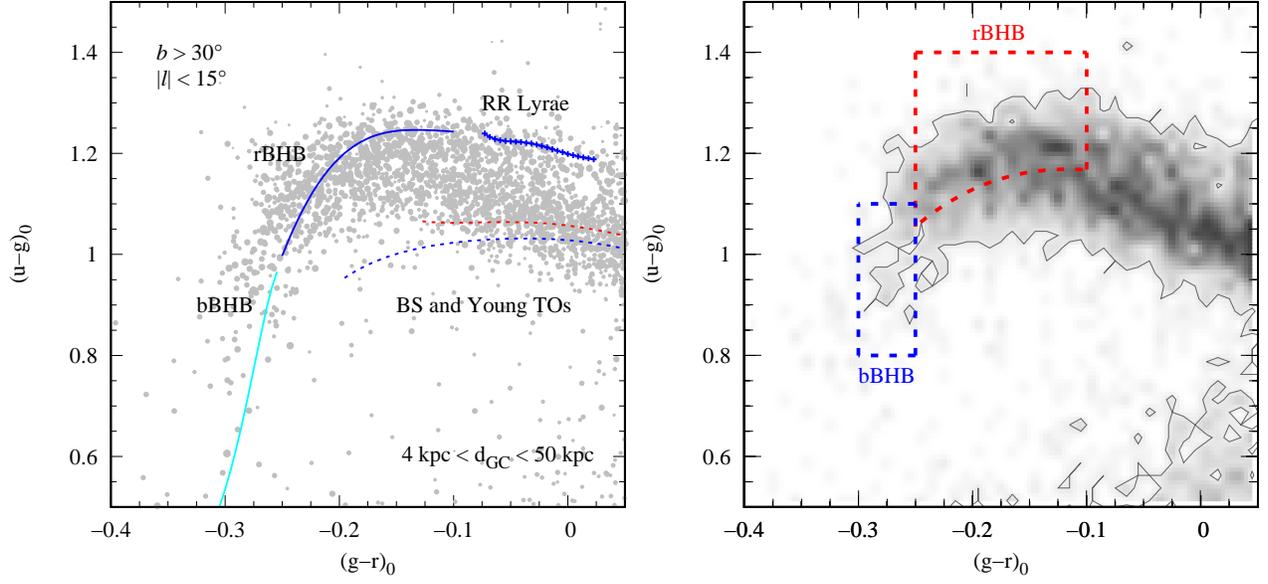}
\caption{Left: volume-limited halo stars selected from SDSS DR14 \emph{PhotoPrimary}.
The position of different evolutionary stages of stars in the $(u-g)_0$ and $(g-r)_0$ diagram are indicated based on HB stars presented in the bottom panels of Figure~\ref{f1}.
Blue solid, cyan solid, and blue line with crosses indicate rBHB, bBHB, and RR Lyrae stars, respectively.
Blue and red dashed lines for BSs and young turn-off stars are predicted by simple stellar population models of ${\rm [Fe/H]}=-1.5$ and $-1.0$.
Right: the density map of halo stars and selection boxes of rBHB and bBHB stars in the color--color diagram.
\label{f2}}
\end{figure}

We use the color--color diagram (CCD), which is not affected by the distance, to select a clean sample of stars in the same evolutionary stages, i.e., rBHBs and bBHBs, in the halo.
Figure~\ref{f2} shows halo stars selected from the SDSS DR14 in the $(u-g)_0$ vs. $(g-r)_0$ plane.
We overplot models of synthetic HB stars as well as young turn-off stars on top of the selected halo stars to show how rBHBs and bBHBs are distributed in the CCD.
The position of these model HB stars on the CCD are indicated as solid lines with blue and cyan colors for rBHBs and bBHBs, respectively.
The adopted metallicities for HB stars in the diagram are the same as models in the bottom panels of Figure~\ref{f1}.
In order to show where blue stragglers (BSs) are placed in the CCD, we provide the dashed lines of young main-sequence turn-off points which cross the lower part of the area.
Blue and red dashed lines are for the metallicities of ${\rm [Fe/H]=-1.5}$ and $-1.0$, respectively.
rBHB stars and some RR Lyraes from metal-poor populations are placed in the curved hook-shaped area in the CCD and sweep the upper edge of the hook-shape.
We intend to select rBHBs and bBHBs simultaneously, so we narrowed down our sample using the guidelines of our population model.
Among the selected sample, we use 5,942 stars that satisfy our criteria for the analysis in Section 3.
The final selection boxes with all samples in the density map are presented in the right panel of Figure~\ref{f2}.
The high-density regions in the CCD are shown as dark black colors.

\section{THE FRACTION OF THE SECOND-GENERATION POPULATIONS IN THE INNER HALO}
\label{newfindings}

\begin{figure}
\centering
\includegraphics[angle=-90,scale=0.65]{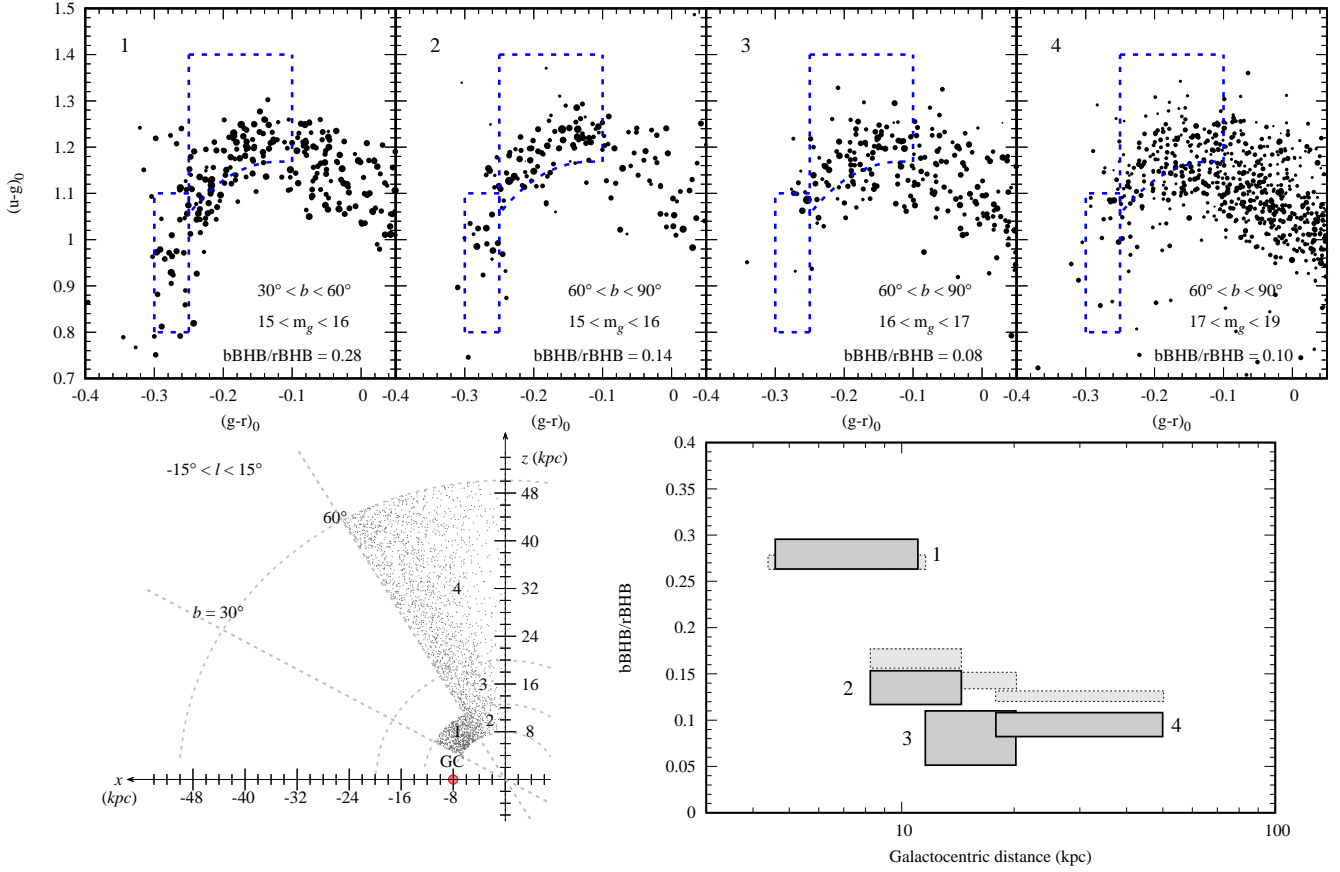}
\caption{(\emph{Upper panels}) The CCD for measurements of the number ratio between rBHBs and bBHBs in the Milky Way halo with respect to the Galactocentric distance.
The point size of stars {is inversely proportional} to the photometric error.
The rBHB magnitude (${\rm M}_g \sim 0.45$) from 15 to 19 corresponds to the Galactocentric distance from 4.0 to 50.0 kpc.
The number of stars within the bBHB selection box is drastically reduced as the Galactocentric distance increases.
(\emph{Bottom left}) The volume-limited halo stars in the heliocentric plane.
The red dot marks the center of the Milky Way.
The denoted numbers on the plot indicate the location of the samples presented above.
(\emph{Bottom right}) The trend of the bBHB-to-rBHB ratio with the Galactocentric distance.
{Dark and pale grey boxes indicate the cases of $\left|l\right|<15^\circ$ and $\left|l\right|<30^\circ$, respectively.
For presentation purposes, a slight extension in the Galactocentric distance was introduced for the broader $l$ case in the 1 region.}
The width of lines indicates the relative errors caused by uncertainty in the number of stars within the selection box. 
{Even with a larger sample, the result shows an almost similar trend to the narrow $l$ case.}
\label{f3}}
\end{figure}

Figure~\ref{f3} shows the selection boxes of rBHBs and bBHBs with respect to the magnitude bin.
The bottom left panel shows the location of our volume-limited samples in the heliocentric coordinate.
To analyze the population shift with respect to the Galactocentric distance, we divided halo stars into four regions, and those regions are indicated by numbers in the plot.
As the magnitude and the Galactocentric distance increase, stars in the bBHB selection box gradually decrease.
Following the bBHB\slash rBHB selection boxes, the fraction of bBHB-to-rBHB decreases with the increasing Galactocentric distance from 1 to 4 region, implying that the fraction of bBHBs increases with the decreasing Galactocentric distance.
The bottom right panel of Figure~\ref{f3} demonstrates this trend with respect to the Galactocentric distance.
{There exists an apparent contrast in bBHB/rBHB ratios between the region inside 10 kpc and the region outside.}
This is consistent with earlier studies based on B-V colors \citep[e.g.,][]{1991ApJ...375..121P} and spectroscopic selection of BHB stars \citep{2015ApJ...813L..16S}.

This result is particularly interesting because it is well established that metallicity of stellar components of Galactic halo increases as Galactocentric distance decreases ~\citep[e.g.,][]{1994AJ....108.1016L, 2015MNRAS.446.2251T}, which in turn result in more red HBs in the inner halo.
Also, the age estimations of Galactic inner halo in the literature suggest around $\sim 11.5$~Gyr \citep[e.g.,][]{2012Natur.486...90K, 2016NatPh..12.1170C}.
{As presented in Figure~4 of \citet{2017ApJ...842...91C}, to reproduce bBHB at halo metallicity without assuming He-enhancement, the age of the population would be at least 13 Gyr.
Therefore, the age referred to in the literature} is not old enough to reproduce BHBs.
Moreover, the age structure of Galactic halo stars shows no radial age gradient \citep[e.g.,][]{2011A&A...533A..59J, 2014MNRAS.445.2575H} or 1~Gyr older ages with the decreasing Galactocentric distance ~\citep[e.g.,][]{2015ApJ...813L..16S, 2016NatPh..12.1170C}, and neither of these is not enough to change the morphology of HBs from red to blue type at given inner halo metallicity ($\rm {[Fe/H]}\sim -1.4$ see \citealt{2015MNRAS.446.2251T}). 
According to these, the general trend of halo populations would naturally produce more red HB stars in the central region of the Milky Way ~\citep[see, ][]{2013ApJS..204....3C}.
In contradiction to the prediction based on stellar populations reproducing red HB stars at a given mean metallicity of inner halo field, our results show the opposite trend.

In order to explain the increasing fraction of bBHB stars in the Galactic inner halo\footnote{The age and CNO abundance also affect the HB morphology. However, considering the increasing N abundance \citep[e.g.,][]{2011A&A...534A.136M, 2017MNRAS.465..501S} and small age difference with decreasing Galactocentric distance, bBHBs are most likely originated from He-rich populations.}, we need to adopt He-rich populations usually observed in GCs.
Since He-rich stars evolve faster than normal He stars, the He-rich stellar populations originated from GCs could reproduce more bBHB stars at younger ages with a high metallicity of inner halo ~\citep{2011ApJ...740L..45C, 2017ApJ...842...91C}.
Figure~\ref{f1} displays this effect by showing the population synthesis model of simple stellar populations under the different assumptions on the initial He abundance.
BHBs are naturally reproduced even at the intermediate metallicity of ${\rm [Fe/H]} \simeq -0.7$.

\citet{2016MNRAS.456L...1C} used BSs and BHBs as proxies for FG and SG populations, respectively, and explained that the low BHB-to-BS ratio in the outer halo of the Milky Way is caused by the preferential removal of the normal He FG stars from proto-GCs while He-rich SG stars remain inside of the proto-GCs.
If this is universal regardless of the Galactocentric distance of the proto-GCs, the fraction between FG and SG, which can be estimated using rBHBs and bBHBs, should remain constant regardless of the Galactocentric distance. 
However, our result shows that the preferential disruption of GC-like systems in the inner halo regime of the Milky Way halo becomes ineffective as the Galactocentric distance decreases but rather the disruption of both FG and SG stars is preferred due to the stronger tidal force of the Milky Way.

The result may lead us to a comprehensive interpretation of the early formation of the Milky Way.
The contribution of disrupted proto-GCs to the Milky Way halo took place globally during the formation of Milky Way bulge and halo.
The preferential removal of the FG stars within the proto-GCs is a function of the Galactocentric distance, and this becomes stronger as the proto-GCs reside in the outer part of the Milky Way halo.
However, for the proto-GCs located in the inner part of the Milky Way, both FG and SG populations are affected by the disruption, leading the increasing SG population in the field with the decreasing Galactocentric distance.
In this sense, the double red clump phenomenon of the Milky Way bulge is the natural outcome of effective removal of the He-rich SG stars from thier host GCs.

\section{Dynamical evolution models of PROTO-GLOBULAR CLUSTERS}
\label{dynamicalevolutionmodelsofproto-gcswithrespecttothegalactocentricdistance}

The observed trend can be explained by a couple of galaxy formation scenarios. One of such possibilities includes a gradual age shift \citep{2015ApJ...813L..16S, 2016NatPh..12.1170C}. 
In this Letter, we propose that it can also be plausibly explained by disruption of massive GCs by tidal force, and subsequent ejection of He-enriched SG stars with systematically bluer colors.

We use our C implementation of the symplectic, time-reversible second order integrator introduced by \citet{2010AJ....139..803Q} to compute the orbits of stars in a Plummer potential (representing the GC) in a circular orbit around a point-mass galaxy (representing the Milky Way). The forces acting on each star include tides and the Coriolis force, which, being velocity-dependent cannot be treated correctly with simpler integration schemes such as standard leapfrog ~\citep{2011MNRAS.415.3168R}.

We generate stars with an initially multivariate normal distribution with standard deviation equal to $a/\sqrt{3}$ along the $x$, $y$, $z$ coordinates, where $a$ is the Plummer scale radius, and equal to $v_c/\sqrt{3}$ along the $v_x$, $v_y$, $v_z$ where $v_c$ is the equilibrium circular velocity at the scale radius. We then tag as SG the stars located within distance $ka$ from the GC center, where $k$ is allowed to vary over different simulations from $0.4$ to $1$ corresponding to an initial SG fraction between $\approx 10\%$ and $60\%$. The orbits are evolved from these initial conditions for $1000$ crossing times and considered escapers if their final position {is} further than $100 a$ from the center of the GC. Over our set of simulations, we change the ratio of the revolution time (around the Galactic center) to the crossing time by several orders of magnitude. This corresponds to changing the ratio of the Plummer scale radius to the Galactocentric distance.

\begin{figure}
\centering
\includegraphics[angle=0,scale=0.8]{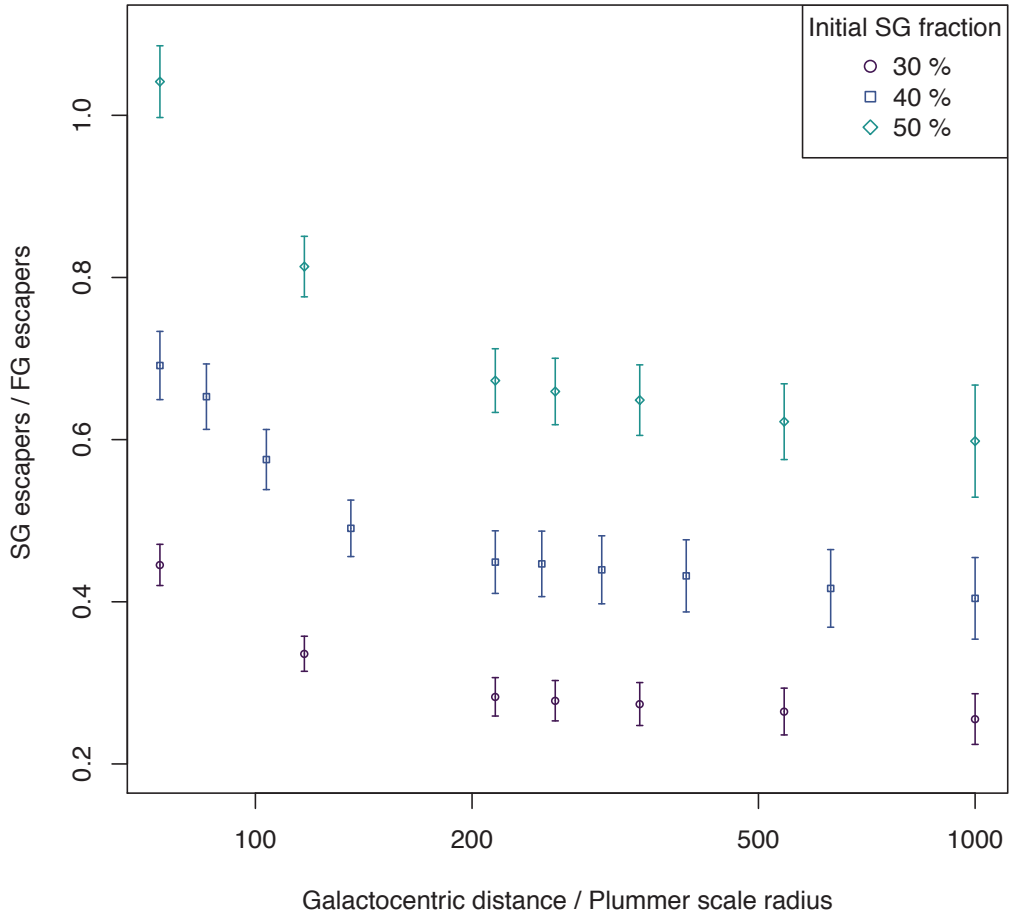}
\caption{{The ratio between SG and FG stars} among the escapers from globular cluster simulations as a function of the Galactocentric radius. Each color corresponds to a different initial fraction of SG stars. At small Galactocentric radii, cluster disruption is almost complete, and the resulting fraction of SG escapers is the same as the initial fraction. As the Galactocentric radius increases, the outer regions of the simulated GC are preferentially stripped, leading to a decreased fraction of SG stars among the escapers.\label{fig4}}
\end{figure}

We find two expected trends in the results of our simulations: first, the fraction of escapers at the end of the simulation is higher in GCs that are nearer to the Galactic center with respect to those that are further away from it. This is clearly a direct effect of tidal stripping. Second, we find that SG stars that are confined within a sphere around the center in the initial conditions of our simulations are less likely to be escapers than the average star. This latter result holds at any Galactocentric distance but is more pronounced far from the Galactic center, where the Hill sphere of the cluster is larger, and stars located centrally are even more unlikely to escape. In Figure~\ref{fig4}, we plot the fraction of SG escapers over {FG} escapers as a function of Galactocentric distance. Simulations with different initial ratios of SG stars over the total number of stars are shown in different colors. The behavior is essentially the same, with almost full stripping at low Galactocentric radii corresponding to a fraction of SG
stars among the escapers equal to the initial fraction. As Galactocentric distance increases, only the stars that were initially unbound and the outermost parts of the GC are stripped, so that comparatively fewer SG stars escape.

\section{Discussion and conclusion}
\label{discussion}

We have found a population gradient in the inner halo of the Milky Way as shown by the decreasing fraction of bBHB-to-rBHB stars with increasing Galactocentric distance based on SDSS photometry.
He-rich SG stellar populations originated from proto-GCs can explain this trend if the fraction of SG tidal stripping increased as Galactocentric distance decreases.
We provide simplified models of two-generation GC evolution in a tidal field, showing that this is indeed the case, at least for dynamically unrelaxed GCs where the SG stars are mostly confined to the GC center.

This suggests that, rather than the contribution of dwarf galaxies only with red HBs, a significant fraction of the halo stellar population is originated from GCs, which built up the Milky Way inner halo during its early formation stages. 
The contribution of SG populations in proto-GCs to the halo assembly depends on the Galactocentric distance of the proto-GCs.
{\citet{2016ApJ...825..146M} already began to discover CN-strong SG stars in the halo, and many more discoveries will be followed based on larger spectroscopic data sets, shortly.}

The simulations we use to reproduce the overall behavior of SG stripping with the Galactocentric radius are greatly simplified.
The most important ingredient which is absent in our models but might affect their outcome is collisional dynamical relaxation, which may lead to a mixing of the SG and FG stars on the relaxation timescale of the cluster \citep[see, e.g.,][]{2013MNRAS.429.1913V}.
We expect this mixing to increase the fraction of SG stars that become tidally stripped, but extensive, direct N-body simulations are required to estimate its full effect (Di Carlo et al. in prep.).

Our discovery leads to fundamental questions about how the Milky Way has formed because the mass fraction of presently remaining GCs seems too small to be the primary building blocks of the Milky Way.
Additional building blocks are still needed to explain the mass of the Milky Way halo and bulge (excluding the disk and bar components).
The recent discovery of clumpy disk galaxies at high redshift ~\citep[e.g.,][]{2017ApJ...836L..22D, 2018NatAs...2...76C} may shed light on the formation history of our Milky Way. 
If these clumps contain $10$ to $100$ proto-GCs inside of them, the total mass of several clumps observed in high-z galaxies is comparable to the mass of galaxy itself.
Inside of these clumps, proto-GCs are provided and they would have experienced chemical evolution similar to the GCs now we observe.
Through this mechanism, the Galactic bulge and halo can be created ~\citep{2008ApJ...688...67E}.
In this way, the significant fraction of the bulge and halo stellar mass can be explained by the contribution of proto-GCs in clumps, also without conflicting with the chemical enrichment scenario of ~\citet{2018ApJ...869...35K}, which can reproduce multiple generations of stellar populations without losing enriched gas from the previous generation stars.

If our hypothesis is correct, we would expect to observe different FG and SG ratios in the present-day GCs with respect to their Galactocentric distance.
However, as shown in \citet{2017MNRAS.464.3636M}, GCs with multiple populations do not show such trends significantly as a function of Galactocentric distance.
This might be explained by the very low surviving rate of proto-GCs, and this makes it hard to find trends in the present GCs with different Galactocentric distances.

Stars in other building block candidates such as dwarf galaxies, mostly consist of FG populations, show slightly different abundance patterns compared to the Galactic inner halo ~\citep{2003AJ....125..707T}.
This implies that Galactic inner halo could be less affected by the satellite dwarfs observed now\footnote{Other competing ideas suggest that massive dwarf galaxies built the inner halo \citep[e.g.,][]{2017A&A...608A.145B, 2018ApJ...862L...1D}, which must have experienced stronger dynamical friction. Their masses are also large enough that they could contain a large number of $\alpha$-enhanced stars originated from asymptotic giant branch stars.}.
The closest example M31 halo also has enhanced $\alpha$-elements which are similar to the abundance patterns of the Milky Way halo ~\citep{2014ApJ...797L...2V}.
Proto-GCs are enriched in $\alpha$-elements fed by SNe type II and they were disrupted to form the halo of the Milky Way rather than present-day dwarf satellites.
This may be yet another manifestation that the primary building blocks of the inner halo are proto-GCs.

Interestingly, recent abundance analyses of early-type galaxies support our predictions of the increasing SG population with the decreasing Galactocentric distance. As shown e.g., in \citet{2017MNRAS.464.3636M} and generally regarded as well established, the enhancement of several elements, such as N and Na, would usually accompany He enhancement. \citet{2017ApJ...841...68V} (Figure 10) directly confirmed our hypothesis of the galaxy formation by showing enhanced Na abundance and depleted O abundance with decreasing radius of early-type galaxies, which are the same behavior of elements originated from GCs. If the dominant fraction of SGs causes this result, the assertion related to the bottom-heavy initial mass function, usually suggested and supported by strong Na absorptions \citep[e.g.,][]{2010Natur.468..940V}, is not valid, but the effect of SG stars originated from GCs is more viable.
We will discuss this issue in the upcoming paper regarding the initial mass function of early-type galaxies (Chung et al. in prep.).

\section*{Acknowledgments}
{We thank the referee for helpful suggestions.} 
C.C., S.-J.Y., D.A., and Y.-W.L. acknowledge support provided by the National Research Foundation (NRF) of Korea to the Center for Galaxy Evolution Research (No. 2017R1A2B3002919, and 2017R1A5A1070354).
S.-J.Y. acknowledges support from the Mid-career Researcher Program
(No. 2019R1A2C3006242).
M.P. acknowledges financial support from the European Unions Horizon 2020 research and innovation programme under the Marie Sklodowska-Curie grant agreement No. 664931. 
U.N.D.C. acknowledges financial support from Universit\`a degli studi dell'Insubria through a Cycle 33rd PhD grant.

\end{document}